\documentclass[aps,prb,reprint,superscriptaddress]{revtex4-2}

\usepackage{graphicx}
\usepackage{amsmath}
\usepackage{amssymb}

\DeclareMathOperator{\Tr}{Tr}

\begin{document}

\title{Mobility edges through inverted quantum many-body scarring}

\author{N. S. Srivatsa}
\affiliation{Department of Physics, King’s College London, Strand WC2R 2LS, UK}
\affiliation{School of Physics and Astronomy, University of Birmingham, Birmingham, B15 2TT, UK}
\affiliation{Max-Planck-Institut f\"ur Physik komplexer Systeme, D-01187 Dresden, Germany}

\author{Hadi Yarloo}
\affiliation{Department of Physics and Astronomy, Aarhus University, DK-8000 Aarhus C, Denmark}

\author{Roderich Moessner}
\affiliation{Max-Planck-Institut f\"ur Physik komplexer Systeme, D-01187 Dresden, Germany}

\author{Anne E. B. Nielsen}
\affiliation{Department of Physics and Astronomy, Aarhus University, DK-8000 Aarhus C, Denmark}

\begin{abstract}
We show that the rainbow state, which has volume law entanglement entropy for most choices of bipartitions, can be embedded in a many-body localized spectrum. For a broad range of disorder strengths in the resulting model, we numerically find a narrow window of highly entangled states in the spectrum, embedded in a sea of area law entangled states. The construction hence embeds mobility edges in many-body localized systems. This can be thought of as the complement to  many-body scars,  an `inverted quantum many-body scar', providing a further type of setting where the eigenstate thermalization hypothesis is violated.
\end{abstract}

\maketitle

When physical systems thermalize, most of the information about their initial state is lost. In the context of quantum mechanics, thermalization is explained through the eigenstate thermalization hypothesis \cite{Deutsch1991,Srednicki1994}, which essentially states that an eigenstate encodes thermodynamic observables characteristic of its energy density. Settings in which quantum systems violate the eigenstate thermalization hypothesis are presently attracting much attention, both for understanding the foundations of many-body physics, and for utilizing their unusual properties, possibly even to store and control the flow of quantum information \cite{nandkishore2015}.

The entanglement entropy of states in the bulk of the spectrum of thermalizing quantum many-body systems is expected to scale with the volume of the system \cite{Page1993,DAlessio2016}. Strong disorder, however, affects all states in the spectrum through the mechanism of many-body localization (MBL) inducing area-law entanglement in the eigenstates and hence nonthermal behavior \cite{Abanin2019}. MBL turns out to be a fragile phenomenon in the sense that there is no agreement of whether it persists beyond a (possibly very long) prethermal timescale  \cite{Vidmar1,Vidmar2,morningstar2022}. The regime of finite systems and finite time scales is, however, by itself interesting and relevant for current experiments \cite{schreiber2015}.

A weaker violation of the eigenstate thermalization hypothesis occurs in systems with quantum many-body scars \cite{Serbyn2021,Moudgalya2022,chandran2022quantum}. Conventionally, scarred states are weakly entangled with subthermal scaling of the entanglement entropy \cite{Moudgalya2018,Turner2018,Schecter2019,Iadecola2020,Chattopadhyay2020,Moudgalya2021,Yao2022}, and  procedures to embed these special states in the bulk of an otherwise thermal spectrum have been developed \cite{Shiraishi2017}.

These studies raise the  question, whether one can also have the converse situation, namely volume law entangled states embedded in a spectrum of MBL states, which we will refer to as inverted quantum many-body scars. Constructing such a model would lead to a different type of nonthermal system beyond MBL and quantum many-body scars. The construction is also interesting from the point of view of mobility edges in MBL. A mobility edge separates localized from delocalized states as a function of energy density, and its existence in the thermodynamic limit, as a matter of principle, is also in question \cite{basko2006,Mondragon2015,roeck2016,Hsu:2018}.

First steps toward constructing inverted quantum many-body scars were taken in \cite{Srivatsa2020a,Srivatsa2022}, where a critical state with logarithmic scaling of the entanglement entropy was embedded in an MBL spectrum, albeit in a model with a highly non-local Hamiltonian and a state with sub-volume law entanglement. A simpler, but still non-local, Hamiltonian was also proposed, but for that case the embedded state was the ground state or a low-lying excited state.

In this paper, we present a local Hamiltonian that allows us to embed a volume law state inside an MBL spectrum. The volume law state is an exact eigenstate for all disorder realizations and hence remains intact even for strong disorder. We specifically consider the rainbow state, which has volume law entanglement for almost all bipartitions. This state is also referred to as an infinite temperature thermofield state \cite{Cottrell2019}.

In quantum many-body scar models, it is quite common that states with energies close to a scar state also have lower entanglement than the thermal part of the spectrum \cite{Turner2018,biswas2022,russomanno2022}. Here, we similarly find that states in the immediate vicinity of the inverted scar state have higher entropy than the MBL states. The number of high entropy states scales exponentially with system size, but with a small enough exponent that the number of high entropy states has measure zero in the large system limit.

The high entropy states produce mobility edges in the localized spectrum. While mobility edges have been observed numerically at the transition from thermal to MBL behavior in several moderate size systems \cite{Luitz2015}, the mobility edges produced by inverted scars are different, as they occur over a broad range of disorder strengths and are particularly sharp as a function of energy density. These properties may be appealing for experimental investigations and practical utilization.

It is well-known that Anderson localized single-particle spectra of non-interacting systems can contain a few delocalized states, as happens, e.g., in quantum Hall systems, when interactions can be neglected \cite{tong2016}. The model presented here differs from that phenomenon in several ways. First, we are here considering a strongly interacting system, and the delocalized states appear in the middle of the \textit{many-particle} spectrum, while the quantum Hall effects happen at low temperature. Second, the rainbow state is volume law entangled, while the delocalized quantum Hall states are less entangled. Third, the quantum Hall effects are eventually destroyed by strong disorder, while in our case the rainbow state is immune to the added disorder.

The model that we investigate also raises interesting questions from a fundamental perspective. The model is generic, except that the disorder must fulfil a particular mirror symmetry. As long as this symmetry is obeyed, the volume law eigenstate persists for all disorder strengths and system sizes. It is well-known that symmetry can lead to delocalization \cite{Nandkishore2014,Banerjee2016,Potter2016}. The special property here is that the symmetry only produces a narrow window of delocalized states rather than delocalizing the entire spectrum. As is generally the case for MBL systems, our finite size numerics is not capable of judging whether the disorder strength at which the transition to MBL takes place remains finite in the thermodynamic limit. If MBL persists in the thermodynamic limit, we expect our (symmetry protected) mobility edges to do likewise.

Finally, we investigate what happens if the mirror symmetry is broken. For a particular system size and disorder strength, we find the inverted scarring to be quite fragile, disappearing already for an admixture of about 0.3\% of non-symmetric disorder.

\textit{Model}---The starting point for our construction of an inverted quantum many-body scar is the so-called rainbow model \cite{Vitagliano2010,Ramirez2014,Ramirez2015} for a chain of $2N$ sites. Here, we consider the general rainbow Hamiltonian
\begin{equation}\label{Ham}
H=H_1\otimes I-I\otimes H_2+c\,V_{\mathrm{int}}
\end{equation}
proposed in \cite{Langlett2022}. $H_1$ acts on the sites $1$ to $N$, and $H_2$ acts on the sites $N+1$ to $2N$. $H_2=MH_1^*M$, where $M$ is the mirror operation that maps site $i$ into site $2N+1-i$, and the complex conjugation is performed in a chosen product state basis. We shall here consider spin-$1/2$ particles with
\begin{multline}\label{H_1}
H_1=\sum_{i=1}^{N-1}(J_xS_i^xS_{i+1}^x+J_yS_i^yS_{i+1}^y+J_zS_i^zS_{i+1}^z)\\
+\sum_{i=1}^N (h_xS_i^x+h_yS_i^y+w_i S_i^z) +J_p\sum_{i=1}^{N-2}S_i^zS_{i+2}^z
\end{multline}
and choose the basis states to be products of eigenstates of the $S_i^z$ operators. Here, $S_i^a$ are the spin-$1/2$ operators for the spin at site $i$. The terms with strengths $J_x$, $J_y$, and $J_z$ describe spin interactions, the terms with strengths $h_x$, $h_y$, and $w_i$ represent a magnetic field, and we include the next-nearest neighbour term of strength $J_p$ to avoid integrability. We take the interaction term in \eqref{Ham} to be $V_{\textrm{int}}=\vec{S}_N\cdot \vec{S}_{N+1}$.

\begin{figure}
\includegraphics[width=\linewidth]{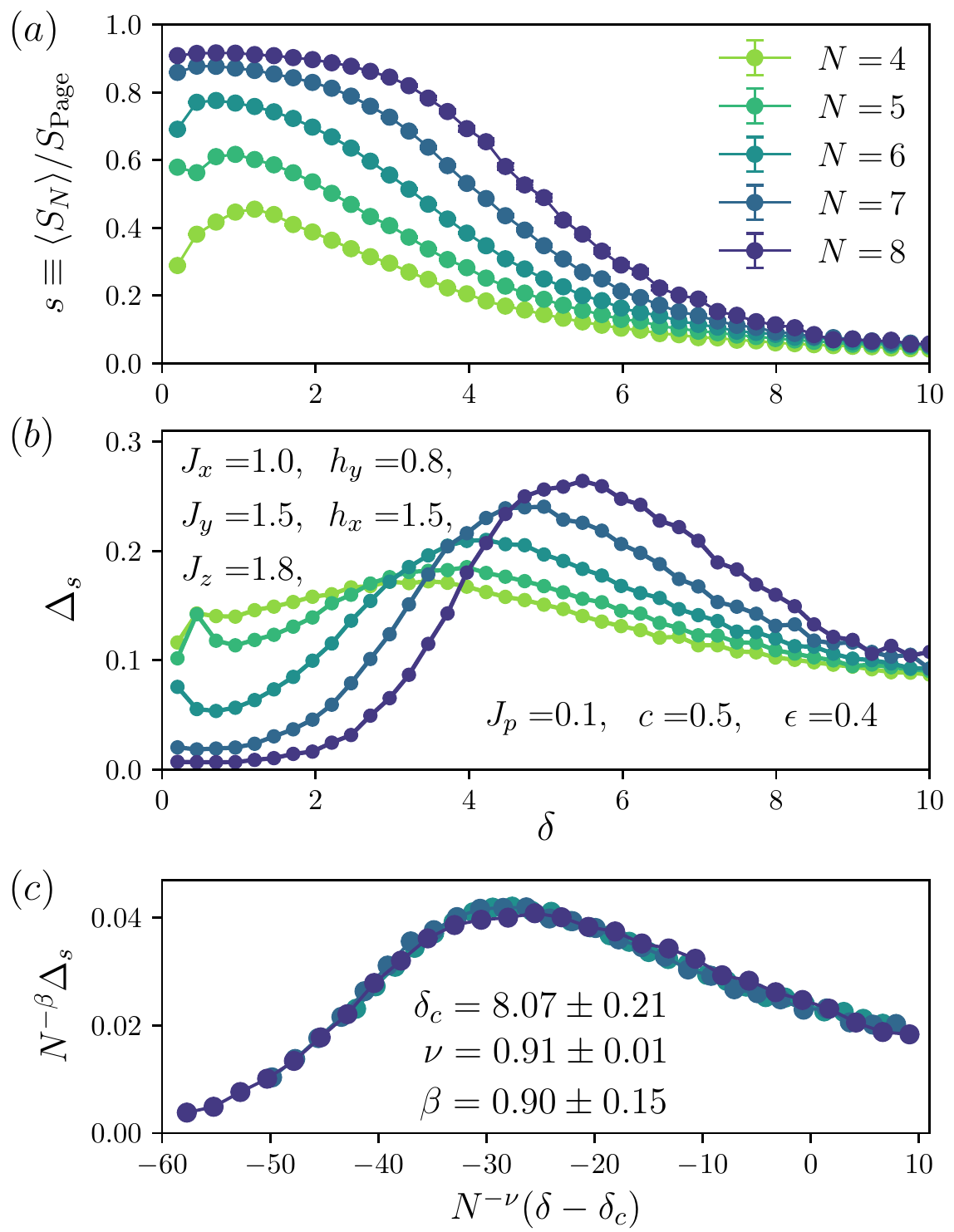}
\caption{(a) Disorder averaged half-chain entanglement entropy divided by the Page \cite{Page1993} value $S_{\mathrm{Page}}=[2N\ln(2)-1]/2$ for the eigenstate closest to the energy density $\epsilon=0.4$ plotted against the disorder strength $\delta$ for different system sizes. The transition from thermal behavior at weak disorder to MBL behavior at strong disorder is seen. The number of disorder realizations is $10^4$ for $N\in\{4,5,6\}$, $5000$ for $N=7$, and $1500$ for $N=8$. (b) The standard deviation of the half-chain entanglement entropy $\Delta_s$ computed for the same set of data shows a peak at the transition point. (c) The finite-size scaling collapse for $\Delta_s$ suggests that the transition happens at $\delta \sim 8$ for large systems.} \label{Fig:entropy}
\end{figure}

It was shown in \cite{Langlett2022} that the rainbow state
\begin{equation}\label{RB-state}
|\psi_{\textrm{RB}}\rangle=
2^{-N/2}\bigotimes_{i=1}^N\left(|{\uparrow},{\uparrow}\rangle_{i,2N+1-i} +|{\downarrow},{\downarrow}\rangle_{i,2N+1-i}\right)
\end{equation}
is an exact eigenstate of $H$ with energy $E_{\textrm{RB}}=c/4$. The rainbow state is a product of Bell states between pairs of spins on opposite halves of the system. The von Neumann entanglement entropy is $\ln(2)$ times the number of Bell pairs that are cut by the chosen bipartition, and hence most choices lead to volume law entanglement \cite{Langlett2022}. The maximal entanglement entropy is achieved for the half-chain bipartition.

We introduce disorder of strength $\delta$ by choosing $w_i$ from a uniform distribution in the interval $[-\delta,\delta]$. The rainbow state is an exact eigenstate independent of the disorder realization. Disorder does, however, affect other states, driving an eigenstate transition from a thermal to an MBL behavior in Hamiltonians with local terms. We show in the following that the disorder indeed many-body localizes the system, except for a set of states near the rainbow state of measure zero. Unless stated otherwise, we take $J_x=1$, $J_y=1.5$, $J_z=1.8$, $h_x = 1.5$, $h_y = 0.8$, $J_p=0.1$, and $c=0.5$ in the computations below. We do not expect the results to be specific to this choice of parameters. The values have been chosen in part to reduce the symmetry of the model and to have the rainbow state close to the middle of the spectrum. The computations for $N\leq6$ are performed by employing full exact diagonalization. For $N>6$, we use the shift-invert spectral transformation, implemented by PETSc~\cite{petsc-user-ref,petsc-efficient}, SLEPc~\cite{Hernandez:2005}, and MUMPS~\cite{MUMPS:2006} to perform Lanczos iteration on the transformed matrix via parallel sparse LU factorization as a direct solver.

\begin{figure}
\includegraphics[width=\linewidth]{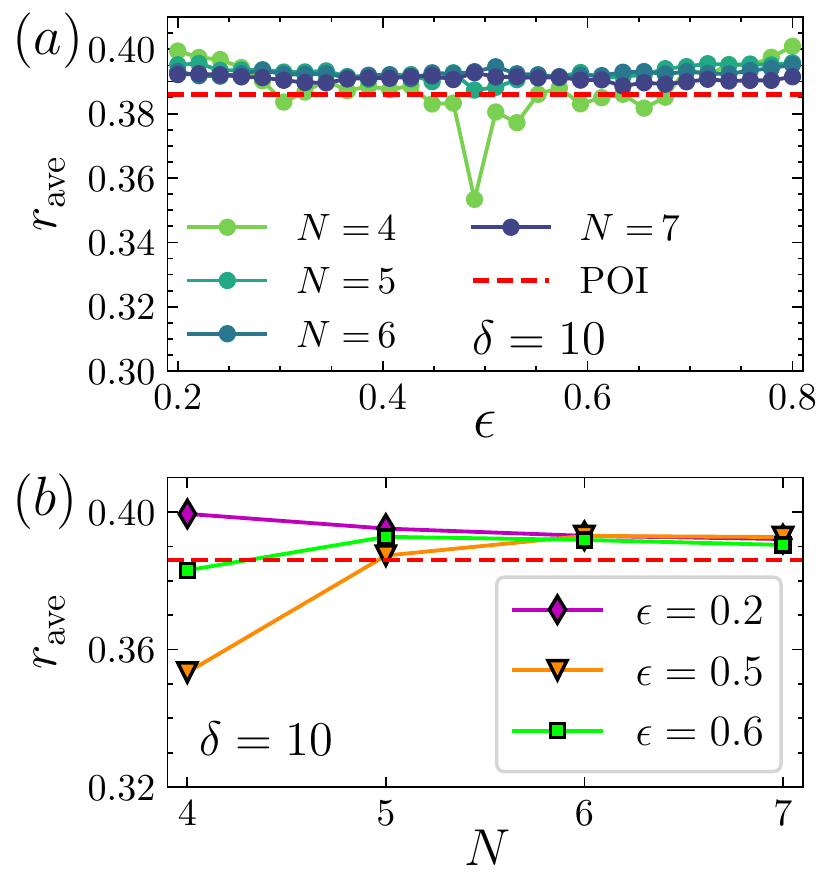}
\caption{(a) Adjacent gap ratio $r_{\mathrm{ave}}$ at strong disorder $\delta=10$ computed for the $13$, $50$, $100$, or $800$ energy levels closest to the considered energy density for $N=4$, $5$, $6$, or $7$, respectively, and averaged over $3000$ disorder realizations. As the system size $2N$ increases, $r_{\mathrm{ave}}$ gets close to the Poisson (POI) value, which signals that the majority of the states in the spectrum are many-body localized. (b) The same data, but plotted as a function of system size for different energy densities.} \label{Fig:rave}
\end{figure}

\textit{Many-body localization}---We first show that the disorder many-body localizes most of the states in the spectrum. We do this by computing the mean and variance of the half-chain entanglement entropy \cite{Kjaell2014} and the level spacing statistics \cite{DAlessio2016}.

We first consider the half-chain von Neumann entanglement entropy $S_N=-\Tr[\rho_N \ln(\rho_N)]$ of an exact eigenstate $|\psi\rangle$ of the system, where $\rho_N=\Tr_{N+1:2N}(|\psi\rangle\langle\psi|)$ is the reduced density matrix obtained after tracing over the spins $N+1$ to $2N$. When averaging the entanglement entropy over disorder realizations, we choose the state with energy density closest to a chosen value in each realization. The energy density is defined as $\epsilon=(E-E^i_{\mathrm{min}}) /(E^i_{\mathrm{max}}-E^i_{\mathrm{min}})$, where $E^i_{\mathrm{min}}$ and $E^i_{\mathrm{max}}$ are the minimum and maximum energies in the spectrum of the $i$th disorder realization and $E$ is the energy of the state $|\psi\rangle$.

In Fig.\ \ref{Fig:entropy}(a), we plot the mean of the entanglement entropy as a function of the disorder parameter $\delta$ for the state closest to the energy density $\epsilon=0.4$. We have chosen this value to consider states close to the middle of the spectrum, while not being too close to the rainbow state, which for most disorder realizations has an energy density close to $0.5$. For weak disorder, the mean entanglement entropy is comparable to the Page value \cite{Page1993}, which signals thermal behavior. For strong disorder, the mean entanglement entropy is independent of system size, which signals MBL. The standard deviation of the entanglement entropy, plotted in Fig.\ \ref{Fig:entropy}(b), shows a peak at the transition point, and the finite-size scaling collapse in Fig.\ \ref{Fig:entropy}(c) suggests that the transition happens at $\delta\sim 8$ for large systems.

\begin{figure}
\includegraphics[width=\linewidth]{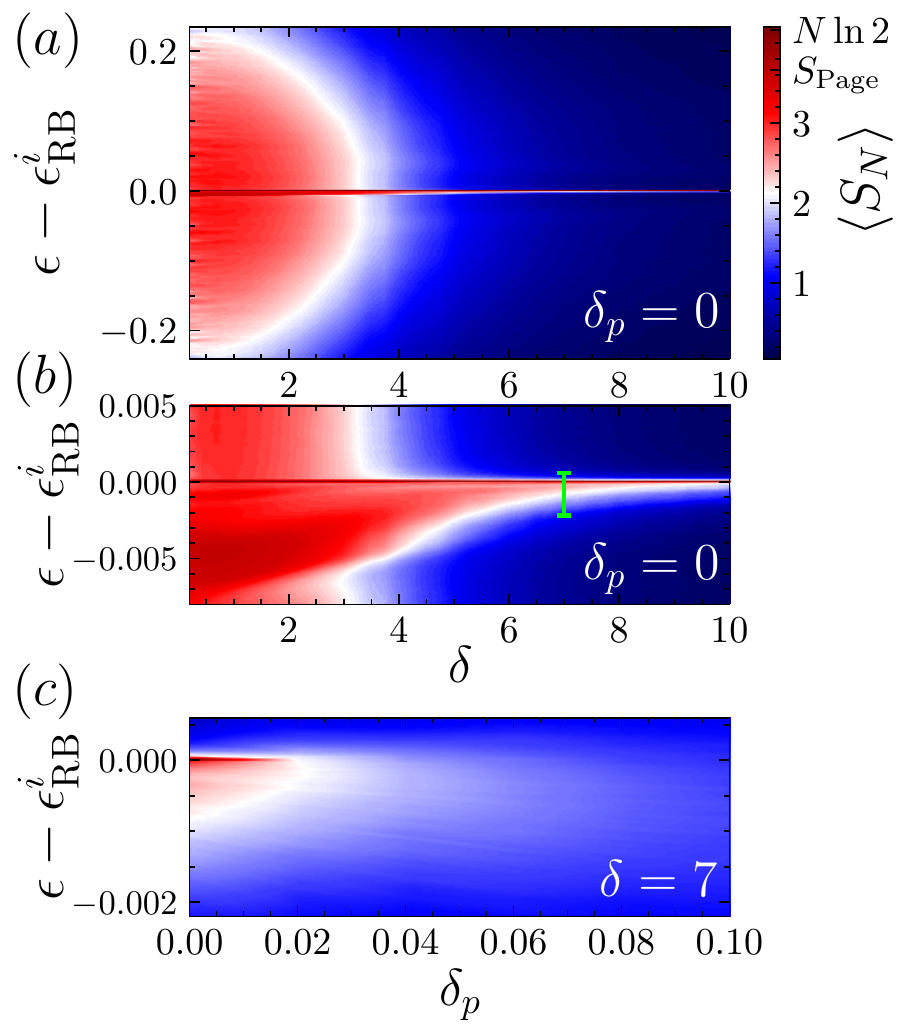}
\caption{(a) The disorder averaged half-chain entanglement entropy $\langle S_N\rangle$ as a function of disorder strength $\delta$ and energy density relative to the rainbow state $\epsilon -\epsilon^i_{\textrm{RB}}$ for $N=6$. We average states with the same value of $n-n^i_{\textrm{RB}}$ over $2000$ disorder realizations, where $n^i_{\textrm{RB}}$ denotes the index of the rainbow state which lies at energy density $\epsilon^i_{\textrm{RB}}=(E_{\textrm{RB}}-E^i_{\min})/(E^i_{\max}-E^i_{\min})$ for the $i$th disorder realization. The dark horizontal line at zero is produced by the rainbow state, and other highly entangled states are seen in its vicinity. Within the strongly disordered regime where the rest of the spectrum is many-body localized, these highly entangled states produce a mobility edge. (b) A zoom of panel (a) showing the band of high entropy states. Most of the high entropy states have energies below the rainbow state, but a few of them are at energies higher than the rainbow state. (c) $\langle S_N\rangle$ for a fixed $\delta=7$ as a function of additional disorder of strength $\delta_p$ on the second half of the chain only.  The high entropy states disappear for $\delta_p\approx\delta/300$. (The symmetric case, $\delta_p=0$, also provides a magnified version of the mobility edges at the cut denoted by the short, vertical, green line in panel (b).)
} \label{Fig:needle}
\end{figure}

The level spacing statistics is another diagnostics to identify whether a system is MBL. Define the energy spacing $\Delta_{n}=E_{n+1}-E_{n}$ and the ratio $r_{n}=\min(\Delta_{n},\Delta_{n+1})/\max(\Delta_{n},\Delta_{n+1})$, where $E_{n}$ is the $n$th energy in the spectrum, and let $r_{\textrm{ave}}$ be the average of $r_n$ over a selected part of the spectrum and over disorder realizations. Arguments from random matrix theory predict that $r_{\textrm{ave}}\approx 0.59$ for thermal states in systems with broken time reversal symmetry, while $r_{\textrm{ave}}\approx 0.386$ in MBL systems.

Figure \ref{Fig:rave} shows $r_{\textrm{ave}}$ as a function of energy density and system size. When the system size increases, the Hilbert space dimension increases, and we hence also average over a larger number of states in the spectrum as detailed in the caption. It is seen that $r_{\textrm{ave}}$ approaches the Poisson value $r_{\textrm{ave}}\approx 0.386$ for large system sizes, which signals that most of the states in the spectrum are many-body localized.

\begin{figure}
\includegraphics[width=\linewidth]{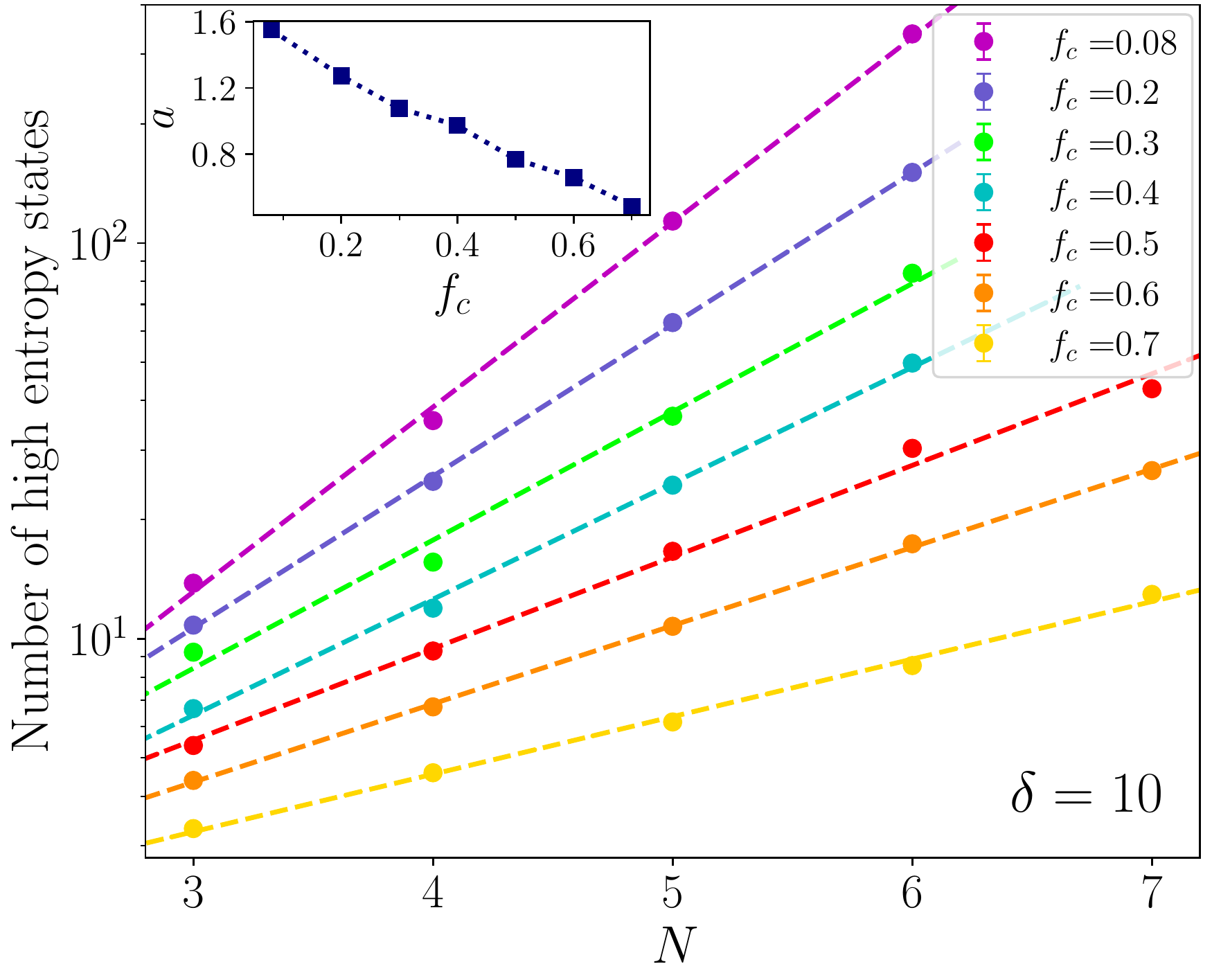}
\caption{Scaling of the number of atypical eigenstates with high entanglement entropy as a function of system size for $\delta=10$. The number of atypical eigenstates is obtained by counting the number of eigenstates with entropy higher than a certain cutoff value, $S_c=f_c\,N \ln(2)$, in each disorder realization, and this number is then averaged over $2000$ disorder realizations for $N=3,4,5,6$ or $1090$ for $N=7$. For all considered $f_c$, the number of atypical eigenstates scales exponentially with $N$, and the dashed lines indicate the best fit with the function $2^{aN+b}$. The inset shows $a$ as a function of $f_c$. Note that $a<2$ for all considered $f_c$, which means that the fraction of atypical states approaches zero for large system sizes.} \label{Fig:scaling}
\end{figure}

\textit{Highly entangled states}---The entanglement entropy of the rainbow state is $N\ln(2)$ for the half-chain bipartition. Since the rainbow state remains unchanged upon introducing disorder, it has a high entropy compared to the many-body localized states, which are area law entangled. We now take a closer look at the behavior of the states in the spectrum that have energies close to the energy of the rainbow state.

Figure \ref{Fig:needle}(a) shows the half-chain entanglement entropy as a function of disorder strength. To probe the states in the vicinity of the rainbow state, we here perform disorder averaging over states that have the same $n-n^{i}_\textrm{RB}$, where $n$ labels the states in the spectrum from lowest to highest energy and $n^{i}_\textrm{RB}$ is the $n$ for the rainbow state for the $i$th disorder realization. The figure shows a narrow band of high entropy states. Crucially, this band is also present for disorder strengths for which the other states in the spectrum are many-body localized. Upon increasing energy, one hence finds a mobility edge followed by an inverted mobility edge. It is interesting to note that the entanglement entropy changes much faster with energy density when crossing the band of high entropy states than it does when crossing the white arc in the left half of the figure that separates the thermal region (red) from the MBL region (blue). We also note that the transition from high to low entropy when crossing the band of high entropy states is particularly sharp as seen in Fig.\ \ref{Fig:needle}(b).

To count the number of high entropy states in the spectrum, we introduce a cutoff $f_c$ and count how many states have an entropy higher than $S_c=f_c N\ln(2)$. This number is plotted as a function of system size for fixed disorder strength and different cutoffs in Fig.\ \ref{Fig:scaling}. It is seen that the number of high entropy states scales exponentially with system size, but the exponent is small enough that the fraction of high entropy states to the total number of states goes to zero in the large system limit.

\textit{Sensitivity to symmetry breaking}---We test the stability of the observed behaviour to a perturbation which distinguishes between the two half-chains in the Hamiltonian \eqref{Ham}, and thereby violates the symmetry underpinning the rainbow state. Concretely, we add further disorder $-w_i+\chi_i$ to the second half of the chain, i.e.\ sites $2N+1-i$, where $\chi_i$ is uniformly distributed within $[-\delta_p,\delta_p]$. We find, for $N=6$ and $\delta=7$ (Fig.\ \ref{Fig:needle}(b)), that high entropy states are formed for $\delta_p\alt0.02$, which is about $0.3\%$ of $\delta$.

\textit{Conclusion}---We have demonstrated a scenario in which an inverted quantum many-body scar with volume law scaling of entanglement entropy is embedded in a spectrum of many-body localized states. The construction does not depend on the microscopic details of the Hamiltonian, except that a specific symmetry constraint needs to be obeyed. Similarly to many quantum scar models, the states in the vicinity of the scar state have modified entanglement compared to the remainder of the spectrum. The number of high entropy states scales exponentially with system size, but not as fast as the dimension of the Hilbert space. The high entropy states thus form a narrow band in the spectrum, demarcated by sharp mobility edges. From our finite-size numerics, we cannot conclude whether the disorder strength at which the transition to MBL happens remains finite in the thermodynamic limit, but if it does, we expect the mobility edges to also remain. The sharp mobility edges over a broad range of disorder strengths in the finite systems may additionally be appealing for experiments and applications. We have also shown that the symmetry constraint does not need to be exactly obeyed to see inverted quantum many-body scarring.

Multiple exact volume law scars may be built by including further symmetries in the Hamiltonian \cite{Langlett2022} and this can lead to an interesting phase with multiple volume law states within a spectrum of MBL states. It would also be interesting to investigate the dynamics in these multiple inverted scar models and study the late time behavior of some simple initial states.

\begin{acknowledgments}
\textit{Acknowledgments}---This work has been supported by Carlsbergfondet under Grant No.\ CF20-0658, by Danmarks Frie Forskningsfond under Grant No.\ 8049-00074B, by the UKRI Future Leaders Fellowship MR/T040947/1, and by the Deutsche Forschungsgemeinschaft under grants SFB 1143 (project-id 247310070) and the cluster of excellence ct.qmat (EXC 2147, project-id 390858490).
\end{acknowledgments}

\end{document}